# Experimental study on Vortex-Induced Vibration of Cylinder at Subcritical Reynolds Number

Zhen Lyu*, Weiwei Zhang and Jiaqing Kou
School of aeronautics, Northwestern Polytechnical University, Xi'an 710072, China
(*Email address for correspondence: zhen_lyu@mail.nwpu.edu.cn)

**Abstract:** Numerical simulation results in recent years show that vortex-induced vibration (VIV) can occur at a subcritical Reynolds number. And the VIV has been observed numerically at Reynolds numbers as low as Re = 20. The current study presents an experimental evidence for the subcritical VIV of a cylinder. We designed and built a rotating channel that makes it possible to perform VIV experiments at subcritical Reynolds numbers. Based on the rotating channel, two sets of tests were carried out for fixed natural frequency with variable incoming flow speed and fixed incoming flow speed with variable natural frequency. In both sets of experiments, subcritical VIV were observed and the VIV can be observed at a Reynolds number as low as 23, which is close to Mittal's numerical results.

## Nomenclature

$m$: mass of the cylinder

$d$: diameter of the cylinder

l: length of the cylinder

$f_n$: natural frequency of the cylinder

$\zeta$: damping ratio of the cylinder

$f$: oscillation frequency of the cylinder

$h$: oscillatory displacement of the cylinder

$\Omega$: angular speed of rotating channel

$U$: speed of incoming flow

$\rho$: density of water

$\upsilon$: Viscosity coefficient of water

Re: Reynolds number, $Re=Ud/\upsilon$

Ro: Rossby number, $Ro= U/2\Omega L$

$F^*$: dimensionless oscillation frequency

$m^*$: mass ratio, $m^*=4m/(\pi\rho d^2)$

$U^*$: reduced speed, $U^*=U/f_n d$

## 1. Introduction

A classic problem in fluid mechanics is the flow past a cylinder. The flow around a stationary cylinder is widely known to become unstable at $Re_{cr} = 47$, with the periodic von Kármán vortex shedding phenomena [1, 2]. When the cylinder is elastically supported, however, the critical Reynolds number can be reduced even further [3, 4]. Within a given range of incoming speeds, an elastically supported cylinder oscillates freely along the direction perpendicular to the incoming flow, accompanied by periodic vortex shedding. This phenomenon is known as vortex induced vibration (VIV) [3,5,6,7]. Cossu and Morino[4] studied the VIV at subcritical Reynolds number by global stability analysis method. They found that at high mass ratios, the critical Reynolds number at which flow instability occurs is reduced to less than half of 47. Bufoni [8] found that vortex shedding can be triggered by forced low-amplitude transverse vibrations at specific frequencies and amplitudes under subcritical conditions down to Re=25. Mittal and Singh [9] found that the minimum Reynolds number for the occurrence of VIV is 20 at a certain support frequency by global linear stability analysis method. They found that lock-in is observed in all subcritical VIV cases. Zhang et al. [10, 11] investigated the effect of transverse and rotational degrees of freedom on subcritical vortex vibration based on linear stability analysis. They discovered that if these two degrees of freedom are released simultaneously, VIV can occur at a minimum Reynolds number of 18. Subsequently, Kou et al. [12] the stable von Kármán vortex shedding mode exists in a subcritical flow regime but becomes less distinct as the Reynolds number decreases. And this mode vanishes when the Reynolds number is less than 18, which is consistent with the lowest Reynolds number for the occurrence of VIV. Recently, Bourguet [13] conducted numerical simulations for subcritical VIV of a flexible cylinder and found that VIV can occur at a Reynolds number as low as 20. In addition, the VIV characteristics in cross-flow direction are different from those in in-line direction.

The conclusions about subcritical VIV presented above are based on numerical simulations and theoretical analysis; nonetheless, these phenomena have yet to be observed in experiments. Experimental studies [6, 14, 15, 16] on VIV of a cylinder have been conducted at Reynolds numbers larger than Re = 47, mainly because experiments on subcritical vortex vibrations are difficult to design. In this work, we report the results of a series of experiments that we have conducted to provide experimental evidence on the existence of VIV at subcritical Reynolds numbers.

## 2. Experimental setup
### 2.1 Rotating channel

We must first create a steady and uniform flow at a suitably low speed (a few centimeters per second) because the Reynolds number under the test condition is relatively low. Since producing such a low flow speed in a recirculating water channel is very difficult, we designed a rotating channel by referring to the rotating tank [17, 18] in geophysical. The rotating channel is shown in Figure 1 and we built it on December 28, 2020. The main parts of this rotating channel consist of two concentric cylinders with diameters of 500mm and 380mm, respectively, fastened to a transparent circular flat plate at one end. A stepper motor is mounted in the center of the circular plate, allowing the stepper motor to spin the entire channel at speeds ranging from 0 to 0.306rad/s. The speed control of the stepper motor is very accurate, resulting in a very consistent rotation speed of the channel. Before starting the experiment, fill the water channel between the two cylinders with a given amount of water, then turn on the stepper motor to make the rotating channel rotate at a specific speed. The viscous effect aligns the speed of the water in the channel with the rotation speed of the two cylinders after a few minutes, resulting in a steady low-speed flow. The temperature of the water in the channel was strictly controlled at 20.0 °C throughout the experiment to eliminate the effect of temperature on the viscosity coefficient.

We used 2D-PIV to quantify the speed distribution of the flowfield at different cross-sections to assess the quality of the flowfield in the rotating channel. At a rotational speed of $\Omega = 0.102$ rad/s, Figure 2 depicts the instantaneous speed distribution of the horizontal cross section of the flume. The flow speed is rather uniform, as can be seen. Figure 3 shows the turbulence intensity distribution in this state, revealing that turbulence in the center of the water tank is roughly 1%. In addition, the effect of rotation on the flow must be considered. The Rossby number [19, 20], defined as $Ro = U/2\Omega L = R/4d$, is commonly used to characterize the influence of rotation on flow. R is the distance between the model and the channel rotation axis, and d is the diameter of the model. The greater the effect of rotation on the flow, the smaller the Ro number is. While the influence of rotation on the flow is negligible when Ro is less than 1, the effect of rotation on the flow is not negligible when Ro is greater than 1. The rotation effect on the flow can be neglected in present study since Ro is in a range of 29.4-58.8, which is substantially larger than 1.

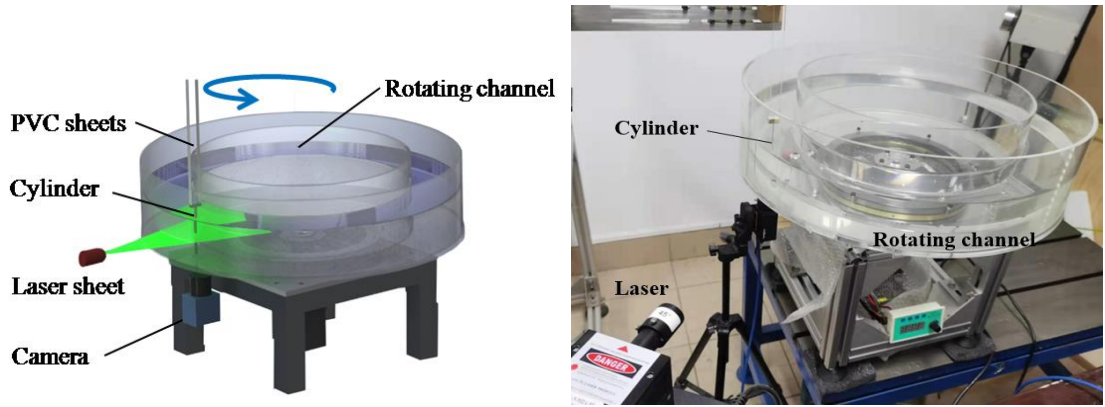

Figure 1. A schematic of the experimental setup

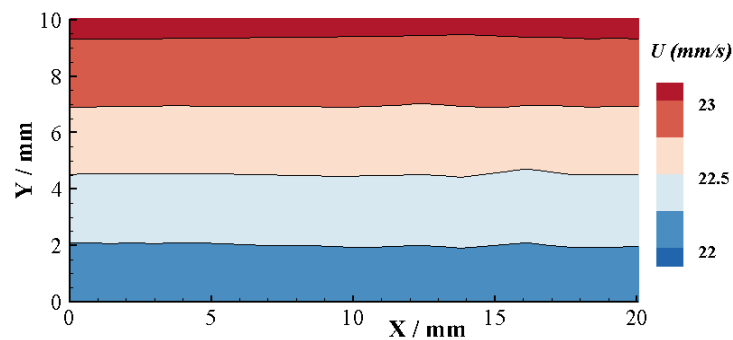

Figure 2. Flowfiled and turbulence distribution at $\Omega$=0.102rad/s

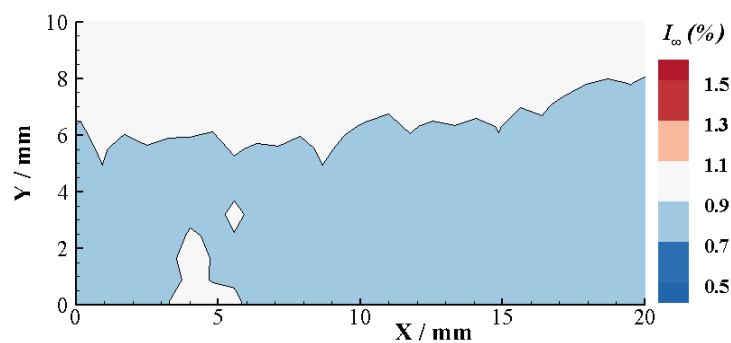

Figure 3. Turbulence intensity distribution at $\Omega$=0.102rad/s

2.2 Test model and measurements

A small aluminum cylinder with a length of 100mm is used as the test model. Depending on the test conditions, the diameter of the cylinder is 1mm or 2mm. The cylinder is held on one end of two parallel flexible PVC sheets and the other end of the PVC sheets are fixed to a clamp. The cross-section size of the PVC sheet is 0.1mm×5mm and the support stiffness can be adjusted by varying the length of the PVC sheets. This mechanism allows the cylinder to vibrate only in the transverse direction perpendicular to the incoming flow, eliminating its degrees of freedom in other directions. The cylinder was placed vertically in the middle of the channel with a submerged height of 90 mm and a distance of 2 mm from the channel floor.

The cylinder's displacement responses were measured with a high-speed camera (Pointgrey GS3-U3-23S6M-C) mounted beneath it, with a displacement measuring resolution of 1.4μm and a sampling rate of 100Hz. The flow around the cylinder was measured by 2D-PIV. Polyamide resion particles with a diameter of 20 μm were used as tracer particles, and a 5 W continuous laser (MGL-N-532A-5W) was used to generate a sheet light source with a thickness of about 1 mm. The laser plane is parallel to the channel floor and 40mm from the bottom of the channel. The PIV images were captured using a Pointgrey GS3-U3-23S6M-C high-speed camera with a resolution of 1920×1200 and a frame rate of 100 fps, resulting in a PIV sampling rate of 100 Hz. The PIV snapshots were post-processed based on the open-source code PIVlab v2.31, using an interrogation window size of 24pixel × 24pixel, an overlap rate of 50%, and a spatial resolution of 0.385mm for speed field measurements.

The natural frequency and damping ratio of the model were measured using a decay test in air. The damping ratios measured in the tests range from 0.006-0.008. Two sets of tests were carried out for fixed natural frequency with variable incoming flow speed and fixed incoming flow speed with variable natural frequency. These tests were conducted in January 2021.

## 3. Results

3.1 Fixed cylinder

To ensure that the experimental setup was correct, we first evaluated the critical Reynolds number of the fixed cylinder. Figure 4 depicts the cylinder's wake distribution at various Reynolds numbers. When the Reynolds number is less than 47, two shear layers can be visible in the wake and do not interact with each other. When the Reynolds number exceeds 47, the vortex shedding of the 2S mode is visible behind the cylinder. This means that the critical Reynolds number of the fixed cylinder in this experiment is 47, which agrees well with the recognized value.

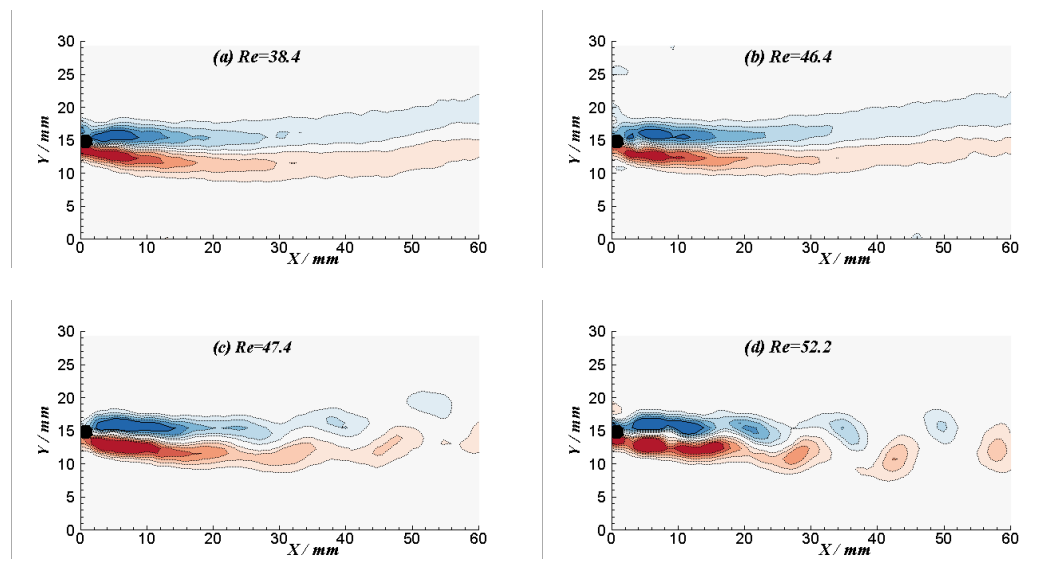

Figure 4. Wake of a fixed cylinder at various Reynolds numbers

3.2 Subcritical VIV at a constant natural frequency

We first carried out experiments with fixed natural frequency varying the incoming flow speed. It is important to note that when changing the incoming flow speed, both Reynolds number and reduced speed change accordingly. In order to achieve higher Reynolds number, the model in this set of tests was an aluminum column of 2 mm diameter. The cylinder has an natural frequency of 1.92 Hz, a damping ratio of 0.0066, and a mass ratio of 4.87. The dimensionless amplitude and oscillation frequency response of the elastically supported cylinder are given in Figure 5*(a),* and it can be seen that the VIV with larger amplitude suddenly starts to appear at U*=6.06. With the increase of reduced speed, the oscillation amplitude increases sharply and reaches the maximum value at U=6.76. When U>6.76, the amplitude decreases gradually with the increase of the reduced speed. The dimensionless oscillation frequencies of VIV at different reduced speeds are shown in Figure 5(b). The dimensionless oscillation frequency is constantly around 1.0, as can be shown. The phenomenon of "lock-in," in which the cylinder's oscillation frequency follows its support frequency, occurs throughout the range of vortex vibrations at low Reynolds numbers. This conclusion is consistent with the results of numerical simulations by Mittal et al. [9]

In addition, the wake of the elastic support cylinder is depicted in Figure 6 at various reduced speeds. At U=6.06, the corresponding Reynolds number is 37.6, which is less than $Re_{cr}$. The cylinder was not oscillating at this time, and there were no vortex shedding in the wake. While U=6.12 and U=6.76, the cylinder oscillates periodically. And Kármán vortex street can still be observed in the wake, even though the Reynolds numbers are still less than $Re_{cr}$. At U=10.23, the amplitude of the VIV is very small, while the Kármán vortex street is clearly visible in the wake. This is because the Reynolds number Re=72.1 at this time, which is larger than the critical Reynolds number.

It is worth noting that when U<7.58, the Reynolds number of the cylinder is less than 47, but VIV is still observed, which verifies the subcritical VIV derived from the numerical simulation. In addition, the subcritical VIV is accompanied by periodic vortex shedding, indicating that the critical vortex shedding Reynolds number of the elastically supported cylinder is less than 47.

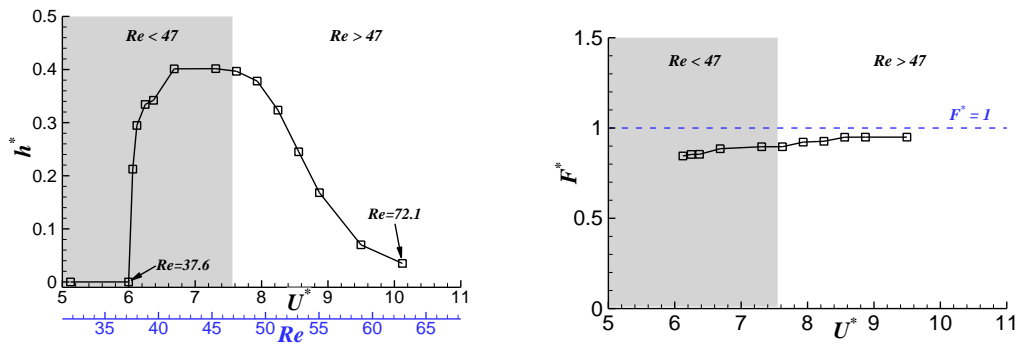

*(a)* Dimensionless amplitude response    *(b)* Dimensionless oscillation frequency response

Figure 5. Amplitude response and frequency response

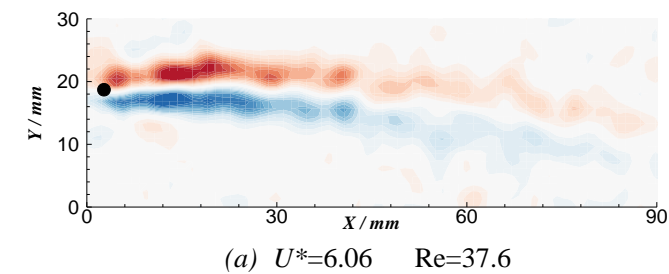

*(a)* U*=6.06    Re=37.6

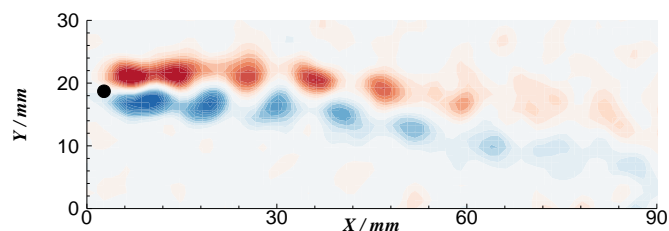

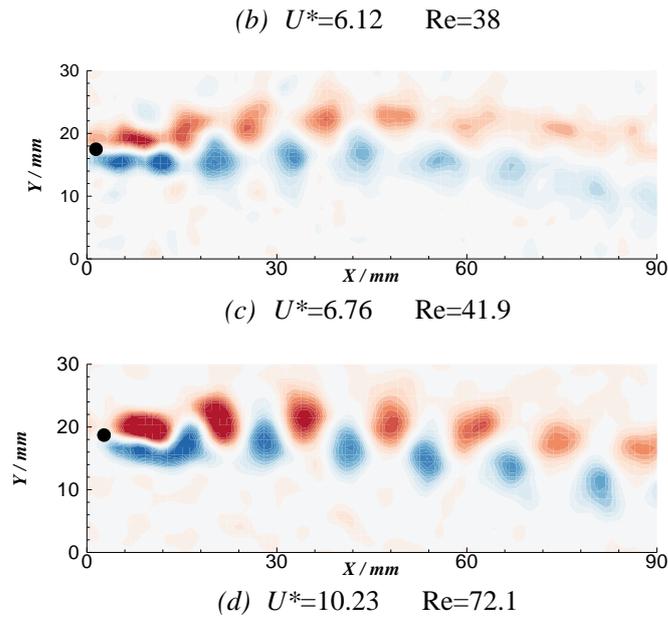

(b) $U^*$=6.12    Re=38

(c) $U^*$=6.76    Re=41.9

(d) $U^*$=10.23    Re=72.1

Figure 6. Wake of the elastically supported cylinder at typical reduced speed

3.3 The lowest Reynolds number of subcritical VIV

To further investigate the lowest Reynolds number for the occurrence of subcritical VIV, we conducted tests at different Reynolds numbers. In order to reduce the Reynolds number, an aluminum cylinder with a diameter of 1 mm was used as the model in this set of tests. The tests are conducted at four different Reynold numbers: 20, 23, 30 and 40. The flow speed is constant at each Reynolds number and the reduced speed is varied by changing the support frequency. The support frequency was varied by changing the length of the PVC sheet in the tests. The additional mass of the PVC sheet decreased as the support frequency increased, resulting in a mass ratio ranging from 7.73 to 9.19.

The amplitude responses of subcritical VIV at four different Reynolds numbers are shown in Figure 7(a). VIV begins to occur in a certain range of reduced speed when the Re≥23. With the increase of Reynolds number, the maximum amplitude of subcritical VIV and the range of VIV also increase gradually. However, for Re=20, VIV was not observed in the test, indicating that the lowest Reynolds number for VIV of a cylinder is 23, which is close to Zhang's [10, 11] conclusion based on linear stability analysis.

In Figure 7(b), the responses of the dimensionless vibration frequency with reduced speed are shown for four different Reynolds numbers. It is worth noting that the dimensionless vibrations frequencies are all around 1 for different Reynolds numbers. This means that the "lock-in" phenomenon is always present in subcritical VIV, which is consistent with Mittal's [9] numerical simulation results.

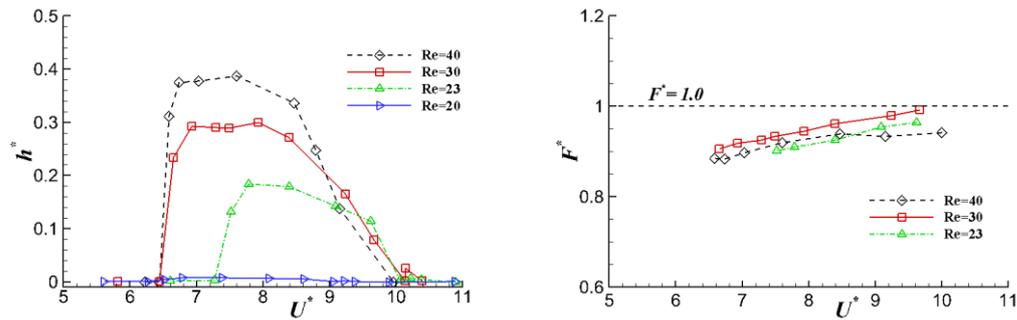

(*a*) Dimensionless amplitude response     (*b*) Dimensionless oscillation frequency response

Figure 7. Amplitude response and frequency response

**4. Conclusion**

In this paper, an experimental study is conducted to verify the existence of the subcritical VIV found by numerical simulation. The lowest Reynolds number for the occurrence of subcritical VIV of the cylinder was found to be 23 in the experiments, which is close to the results of the numerical simulation by Mittal et al. [9] With the increase of Reynolds number, the maximum amplitude of subcritical VIV and the range of VIV also increase gradually, but the vibration frequency of the cylinder is always locked on the support frequency. Additionally, periodic vortex shedding with frequencies corresponding with the vibration frequency are observed downstream of the cylinder during subcritical VIV. This indicates that the elastic support reduces the stability of the flow around the cylinder, making it unstable at Reynolds numbers below 47. This finding is consistent with Zhang's [10, 11] findings based on linear stability analysis.